\documentclass[%
 reprint,
 amsmath,amssymb,
 aps,
]{revtex4-1}

\usepackage{natbib}
\usepackage{graphicx}
\usepackage{dcolumn}
\usepackage{bm}
\usepackage{xcolor}
\usepackage{soul}
\usepackage{todonotes}
\begin{document}

\preprint{APS/123-QED}

\title{Edge transport in InAs and InAs/GaSb quantum wells}

%Comparing trivial InAs and InAs/GaSb edge transport}

\author{Susanne Mueller, Christopher Mittag, Thomas Tschirky, Christophe Charpentier, Werner Wegscheider, Klaus Ensslin, Thomas Ihn}
\affiliation{Solid State Physics Laboratory, ETH Zurich, 8093 Zurich, Switzerland}

\date{\today}

\begin{abstract}

We investigate low-temperature transport through single InAs quantum wells and broken-gap InAs/GaSb double quantum wells. Non-local measurements in the regime beyond bulk pinch-off confirm the presence of edge conduction in InAs quantum wells. The edge resistivity of 1-2\,$\mathrm{k\Omega/\mu m}$ is of the same order of magnitude as edge resistivities measured in the InAs/GaSb double quantum well system. Measurements in tilted magnetic field suggests an anisotropy of the conducting regions at the edges with a larger extent in the plane of the sample than normal to it. Finger gate samples on both material systems shine light on the length dependence of the edge resistance with the intent to unravel the nature of edge conduction in InAs/GaSb coupled quantum wells.

\end{abstract}

\pacs{Valid PACS appear here}

\maketitle

\section{Introduction}

Topological insulators have been predicted \cite{KanePRL2005, BernevigScience2006} to show the quantum spin Hall effect based on dissipationless transport in edge states separated by an insulating bulk. Experimentally such a situation was first realized in an inverted HgTe/(Hg,Cd)Te quantum well by the proper choice of quantum well thickness \cite{KoenigScience2007}. While a non-local measurement proved the existence of edge transport \cite{RothScience2009}, the confirmation of  spin-polarized transport demanded more complex transport experiments \cite{BruneNature2012}. For the coupled quantum well system InAs/GaSb, double gating was predicted to tune density and band alignment independently \cite{LiuPRL2008, QUPRL2015} resulting in a tunable two-dimensional topological insulator. According to theory, conductance in helical edge states is switched on or off when crossing the boundary between the topological and trivial insulator by a proper change of front- and back-gate voltage. 

In a series of pioneering experiments the group of R. R. Du reported evidence for edge modes in inverted InAs/GaSb quantum wells \cite{KnezRRL2011,KnezPRL2014}. In addition, quantized conductance close to charge neutrality was reported \cite{DuPRL2015}. Subsequently, edge conduction was confirmed by a number of groups in the regime of inverted band alignment by non-local transport measurements \cite{SuzukiPRB2013, KnezPRL2014, MuellerPRB2015}, by scanning SQUID \cite{SpantonPRL2014} and via the detection of edge-mode superconductivity \cite{PribiagNatureNanoTech2015}. So far there is no experimental report in the literature directly demonstrating the helical nature of these edge states. Several publications \cite{KnezRRL2011,SuzukiPRB2013, MuellerPRB2015} reported on the relevance of  bulk conduction and limited gate tunability which prohibit to study edge conduction in the full phase diagram and hamper, for example, a detailed survey of the length dependence of the  edge conductance.

Recent findings of edge conduction in the non-inverted regime \cite{NicheleNJP2016, NguyenPRL2016} raised additional questions. The physical origin of these edge states is under debate. In addition, such trivial edge states may possibly co-exist with helical edges in the inverted regime and it is unclear how such edge states of different origin could be distinguished by transport experiments.

Here we report edge transport experiments on InAs quantum wells, having in mind that this material is a constituent of InAs/GaSb double quantum wells. We consistently find edge conduction with a resistivity of 1.3-2.5\,k$\Omega$/$\mu$m with different techniques. Tilted magnetic field measurements suggest an anisotropy of the conducting edge channels with a larger extent in the plane of the sample. Analogous transport experiments in InAs/GaSb devices allow us to study the dependence of conductance on edge length in the inverted regime. 

This paper is structured as follows: After an introduction to wafer material, sample fabrication and measurement setup in section II, we present the results on InAs devices of different geometries in section III. In section IV we then compare to measurements on InAs/GaSb double quantum wells. Measurements in tilted magnetc fields are jointly presented for both material systems in section V, before we finally attempt to find a consistent interpretation of all the presented data and critically discuss the conclusions that can be drawn in section VI.

\section{Wafer material, sample fabrication and measurement setup}

Three different wafers were grown by molecular beam epitaxy. Wafers A and B host a two-dimensional electron gas in an InAs quantum well. Wafer C contains a two-dimensional electron and hole gas in a InAs/GaSb double quantum well. Wafer A is grown on a GaAs substrate and contains a 15\,nm InAs quantum well confined by AlSb barriers. The layer sequence of wafer A was also used for wafer C, differing only by the additional 8\,nm GaSb quantum well on top of the InAs quantum well and by the use of a Ga-source with reduced purity as described in Ref.~\onlinecite{CharpentierAPL2013}. Wafer B is grown on a GaSb substrate and has a layer sequence of Al$_\mathrm{x}$Ga$_\mathrm{1-x}$Sb/InAs/AlSb with a 24\,nm InAs quantum well showing an improved mobility as reported in Ref.~\onlinecite{TschirkyPRB2017}. The information on the epitaxial growth of wafers A, B and C are summarized in Table~I.

\begin{table*}
	\centering
		\begin{tabular}{|l|l|l|l|l|l|}
			\hline
			\textbf{Wafer name} & \textbf{Quantum well material(s)} & \textbf{Well thicknesses} & \textbf{Barrier materials} & \textbf{Substrate} & \textbf{Ga-source purity}\\
			\hline
			\hline
			A & InAs & 15\,nm & AlSb and AlSb & GaAs & high\\
			B & InAs & 24\,nm & Al$_\mathrm{x}$Ga$_\mathrm{1-x}$Sb and AlSb & GaSb & high\\
			\hline
			C & InAs/GaSb & 15\,nm/8\,nm & AlSb and AlSb & GaAs & low\\
			\hline
		\end{tabular}
		\caption{Wafer details including information on layer sequence, substrate wafer and Ga-source purity used during the epitaxial growth.}
\end{table*}

Hall bar structures were patterned by optical lithography combined with wet chemical etching deep into the lower barrier material as described in Ref.~\onlinecite{PalAIP2015}. Ti/Au pads separated from the wafer surface by a 200-nm-thick Si$_\mathrm{3}$N$_\mathrm{4}$ dielectric are used as gates. Ohmic contacts were made by a Au/Ge/Ni eutectic.

If not mentioned otherwise, the measurements were conducted at 1.5\,K. Four-terminal resistance measurements on wafer A and B were performed by applying an ac current of 10\,nA with a frequency of 31\,Hz. On devices from wafer C, four-terminal DC measurements were conducted because of high contact resistances (of the order of $10\,\mathrm{k}\Omega$) in these devices. The Corbino device was measured by applying an alternating voltage and measuring the AC current with an IV-converter.

\begin{figure}
  \includegraphics[width=\columnwidth]{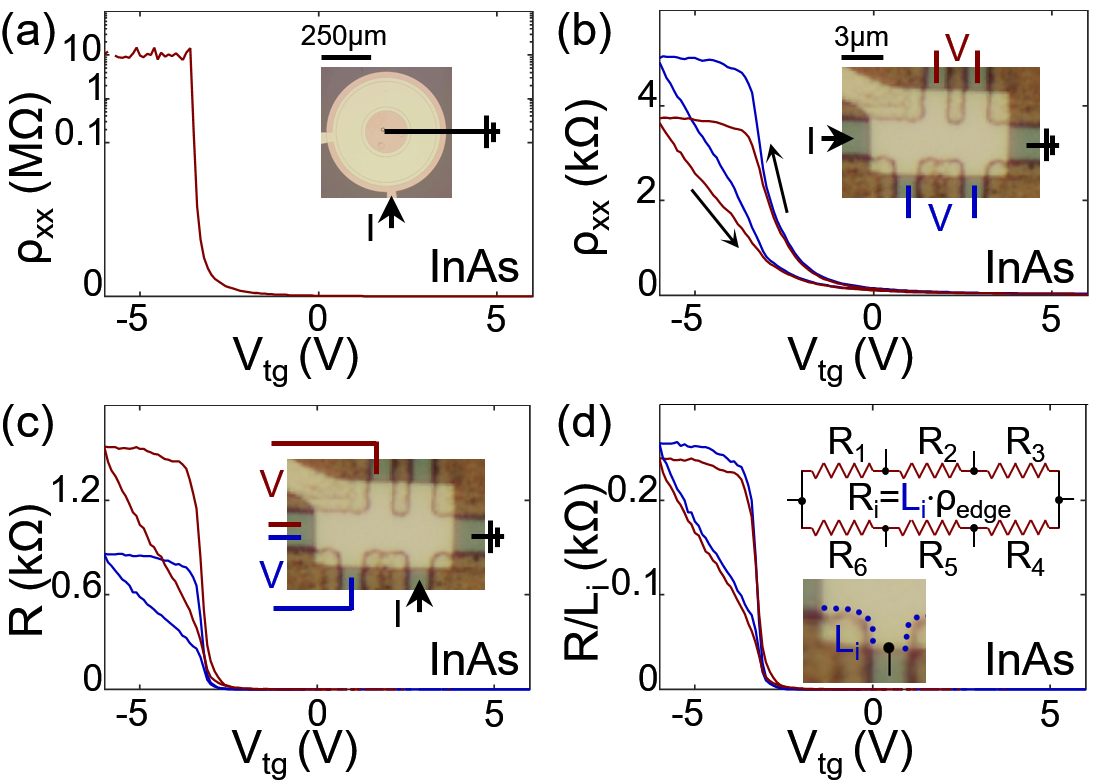}
  \caption{Transport results on InAs two-dimensional electron gases, Wafer A: The longitudinal resistivity $\rho_{xx}$ as a function of top gate voltage $V_\mathrm{tg}$ of a Corbino device (a) and of an asymmetric Hall bar (b). Arrows indicate the gate-voltage sweep-direction. The optical microscope pictures together with the measurement configuration are shown as insets. (c) Finite non-local resistances $R$ measured in the configuration indicated in the inset. (d) Measurement results from (c) normalized by the respective edge segment length $L_\mathrm{i}$ [dotted lines in the lower inset of (d)]. The resistor network model in the upper inset is explained in the text.}
\end{figure}

\section{Transport in $\mbox{InAs}$ devices of different geometries}

%We start by discussing the results obtained on (trivial) InAs 2DEGs. By applying a suitable top gate voltage $V_\mathrm{tg}$ the Fermi energy falls within the band gap of this semiconductor and the 2DEG is depleted. One expects a high bulk resistance in the so-called pinch-off regime. In theory, no edge states are expected.
Figure 1 shows the top-gate ($V_\mathrm{tg}$) dependent resistance at zero magnetic field of different device geometries fabricated on wafer A. The Corbino device [Fig.~1(a), inset] allows us to deduce the bulk resistivity from the measured conductance, in the absence of any edges connecting the two ohmic contacts. For positive gate voltages $V_\mathrm{tg}$ the longitudinal resistivity is $\rho_{xx}\approx30\,\Omega$ indicating that the Fermi energy is deep in the conduction band. Around $V_\mathrm{tg} \approx -2.5\,$V the resistivity $\rho_{xx}$ increases rapidly signaling the depletion of the two-dimensional electron gas. It reaches approximately 10\,M$\Omega$ for $V_\mathrm{tg} < -3.6\,$V, which is the measurement limit due to parasitic cable capacitances in parallel to the sample. Measurements at lower frequencies (not shown) give a lower limit for the bulk resistivity of G$\Omega$s in this insulating regime.

The transport behavior on a Hall bar device fabricated on the same wafer displays different characteristics as shown in Fig.~1(b). The resistivity reaches a maximum of $\rho_{xx}\approx5~\mathrm{k}\Omega$, rather than the 10\,M$\Omega$ expected from the results of the Corbino geometry. This Hall bar device cannot be pinched-off even at gate voltages well below $V_\mathrm{tg} < -4\,$V. As a preliminary observation, we state that an additional conductive channel must be present in a device with edges as compared to the Corbino device without edge contributions. 

The measurements were conducted on the Hall bar presented in the inset of Fig.~1(b). On this device, two longitudinal voltages can be recorded between leads with different separation. Current and voltage probes are indicated in the inset. The two deduced resistivities agree with each other for $V_\mathrm{tg} > -2.8\,$V, but deviate for $V_\mathrm{tg} < -2.8\,$V. This shows that in the low gate-voltage regime the resistance has a geometry dependence different from the usual length/width-scaling in diffusive two-dimensional systems.

In Fig.~1(b) the sweep direction of the top gate $V_\mathrm{tg}$ is indicated with black arrows. The hysteresis, i.e. the difference in resistivities between the two sweep directions, depends strongly on the gate voltage range. Measurements above $V_\mathrm{tg} = -3.4\,$V hardly suffer from hysteresis effects. For down sweeps the resistivity is almost gate voltage independent below $V_\mathrm{tg} < -3.6\,$V, showing a plateau-like resistivity. However, as soon as the sweep direction is reversed, the resistivity drops. The absence of hysteresis in the Corbino device at all gate voltages supports the interpretation that the hysteresis in Hall bar devices is related to the sample edges.
 
Non-local measurements on the same device shown in Fig.~1(c) reveal further insights into the properties of the additional conductance contribution. An exemplary measurement configuration is shown in the inset. At voltages $V_\mathrm{tg} > 0\,$V the non-local voltage is vanishingly small, compatible with bulk-dominated two-dimensional diffusive conduction. Based on the Corbino-results, we expect the bulk to be depleted at gate voltages $V_\mathrm{tg} < -3.6\,$V, where the non-local resistance in Fig.~1(c) appears to be finite. We interpret this finite non-local conductance as evidence for conduction along the sample edge. 

%Pursuing the idea of edge conduction, we normalize the two non-local resistances shown in Fig.~2(c) (red and blues traces) by the respective gated edge lengths $L_\mathrm{i}$ between the measurement contacts (see inset of Fig. 2(d)). The result in Fig.~2(d) demonstrates that the measurements can be scaled on top of each other, which justifies the procedure and the interpretation. In fact all measured non-local resistances scale similarly to an edge resistivity $\rho_\mathrm{edge} =\mathrm{R_{i}}/\mathrm{L_{i}}= 2.5\pm0.2\,\mathrm{k}\Omega/\mu$m.

Edge conduction in these InAs two-dimensional electron gas devices becomes dominant in transport regimes where the bulk resistivity exceeds the edge resistivity. The bulk resistivity can be tuned only below the gate seen as the yellow shaded rectangle in the inset of Fig.~1(c). We normalize the two non-local resistances $R=V/I$ shown in Fig.~1(c) (red and blues traces) by the respective gated edge lengths $L_i$ between the measurement contacts [see inset of Fig. 1(d)] to demonstrate the linear length dependence. This scaling suggests that the sample can be modeled with a resistor network as shown in the upper inset of Fig.~1(d). All possible four-terminal measurement configurations on this device are consistent when scaled with respective edge segment lengths $L_i$ leading to the edge resistivity $\rho_\mathrm{edge} = R_{i}/L_{i} = V/I_{i}L_{i} = 2.5\pm0.2\,\mathrm{k}\Omega/\mu$m (current flow along the edge $I_{i}$ calculated with the model). This consistency confirms in retrospect the resistor network model used for the analysis.

Similar measurements and analysis were conducted on six Hall bar devices fabricated by wet etching \cite{PalAIP2015} on the two different wafers A and B. All measurements agree with the above findings and result in edge resistivities in the range of $\rho_\mathrm{edge} = 1.3-2.5\,\mathrm{k}\Omega/\mu$m. The wafers used have different quantum well thicknesses, are confined by different lower barriers and were grown on different substrates, each requiring a different growth procedure for the buffer. The observed edge resistivity was independent of all of these boundary conditions. Two Hall bars were fabricated with a dry etching technique (also described in \cite{PalAIP2015}). These devices showed edge resistivites of the same order of magnitude as well. 

%The proportionality of the resistance to the edge length was not found. Still, these measurements support the existence of trivial edge conduction in InAs quantum wells, fabricated with state-of-the art techniques.

Edge states, or more generally, conducting surfaces, were already suspected previously to reduce the efficiency of IR detectors based on InAs/GaSb superlattices \cite{PlisLPR2013} and therefore were studied optically. Fermi level pinning in the conduction band of InAs is often mentioned as a possible reason for the enhanced electron density at the edge, especially because the effect is robust against changes of layer sequence and fabrication \cite{MeadPRL1963, TsuiPRL1970, NoguchiPRL1991, OlssonPRL1996}. Others also add the effect of electric field line concentration, present when gates overlap the sample edges. Edge conduction may also be a side effect of sample processing, either after long exposure to air \cite{ChaghiSST2009} or because of conducting Sb residues on the surface after etching \cite{GatzkeSST1998}. Our experiments add to these results that edge conduction can occur in pure InAs quantum well samples and that it may dominate transport if the bulk is insulating. Its physical origin cannot be assessed by the present experiments.

\begin{figure}
  \includegraphics[width=\columnwidth]{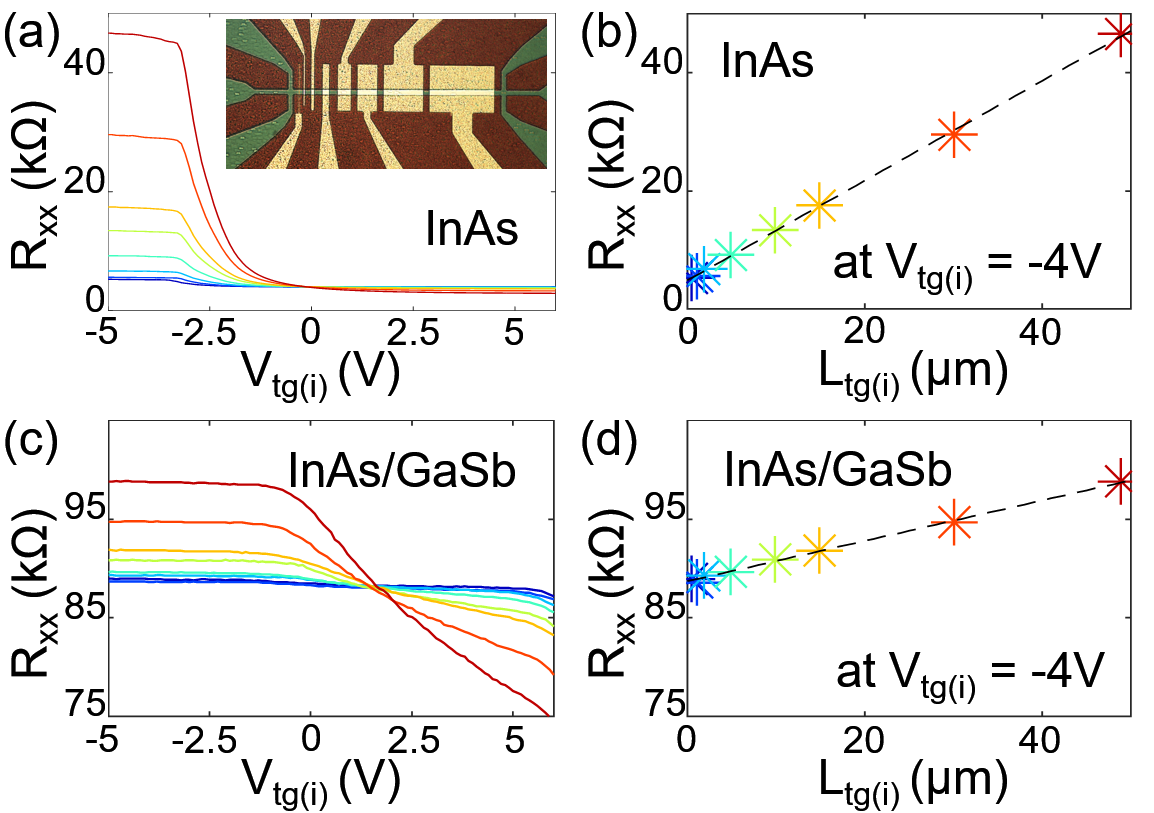}
  \caption{Finger gate samples (optical microscope picture of a device in the inset of (a)) to study the longitudinal resistance $R_{xx}$ as a function of all eight gates with gate lengths reaching $L_\mathrm{tg(i)}=0.5$ to $49\,\mu$m for an InAs two-dimensional electron gas, wafer A (a) and for a coupled quantum well InAs/GaSb, wafer C (c). The longitudinal resistance $R_{xx}$ at $V_\mathrm{tg(i)}=-4\,$V is plotted versus the gate length $L_\mathrm{tg(i)}$ together with a linear fit (black dashed line) in (b) resp. (d).}
\label{finger}
\end{figure}

We continue with an investigation of the length dependence of the edge resistance inspired by the experiments of Nichele {\it et al} \cite{NicheleNJP2016}. The device is shown in the inset of Fig.~2(a). Eight gates, 0.5 to $49\,\mu$m in width, cross a long Hall bar (width $W=4\,\mu$m, length $L=162\,\mu$m). The longitudinal resistance $R_{xx}$ plotted in Fig.~2(a) is the sum of a gate-voltage independent resistance due to the ungated sections of the Hall bar and the resistance caused by the gated section below one of the finger gates biased with $V_\mathrm{tg}^{(i)}$ (note that the constant resistance also depends on the length of the biased gate), i.e.,
\begin{equation}
R_{xx}(V_\mathrm{tg}^{(i)}) =\rho_\mathrm{gated}(V_\mathrm{tg}^{(i)})L_\mathrm{tg}^{(i)} + \rho_\mathrm{ungated}(L_\mathrm{total}-L_\mathrm{tg}^{(i)}).
\label{ldep}
\end{equation}
A systematic length dependence is seen in Fig.~2(a), where only $V_\mathrm{tg}^{(i)}$ down-sweeps are shown for simplicity.

The longitudinal resistance at $V_\mathrm{tg}^{(i)}=-4$\,V is plotted versus the respective gate length $L_\mathrm{tg}^{(i)}$ in Fig.~2(b). The black dashed line is a linear fit to the data according to eq.~\eqref{ldep} using $\rho_\mathrm{ungated}$ and $\rho_\mathrm{gated}$ as fitting parameters, resulting in an edge resistivity $\rho_\mathrm{edge} = 2\rho_\mathrm{gated}(-4\,\mathrm{V}) \approx 1.7~\mathrm{k}\Omega/\mu$m (the factor of 2 accounts for the two edges). The fit line also serves as a guide to the eye to demonstrate the obvious proportionality between longitudinal resistance $R_{xx}$ and the gated edge length.

\section{Comparison to transport in $\mbox{InAs/GaSb}$}

With these insights about InAs in mind we now turn to the discussion of InAs/GaSb double quantum well structures, which contain a hybridized electron-hole system \cite{DrndicAPL1997, NavehAPL1995, MagriPRB1999, CooperPRB1997, NichelePRL2014, NguyenAPL2015}.
%In addition, this system was predicted to exhibit the QSHE. Two counter-propagating, spin-polarized and ballistic edge states are characteristic for this topologically protected regime. The theoretical expectation of the topological regime is a finite, quantized edge resistance and ideally a pinched-off bulk. 
Despite multiple affirmations of edge conduction in the inverted regime of InAs/GaSb \cite{KnezRRL2011,KnezPRL2014,DuPRL2015,SuzukiPRB2013,MuellerPRB2015,SpantonPRL2014,PribiagNatureNanoTech2015}, a careful analysis of the various possible contributions to edge conduction is missing. 

Figures 2(c) and (d) show measurements on an InAs/GaSb finger gate sample (wafer C, same sample dimensions as the InAs sample) analogous to Figs.~2(a) and (b). Also in this device the resistance depends linearly on gate segment length, in agreement with Ref.~\onlinecite{DuPRL2015}. Analogous results could be found on a second device with six gates, 0.5 to 20\,$\mu$m in width, crossing a $2\,\mu\mathrm{m}\times76\,\mu\mathrm{m}$ Hall bar (data not shown). For the finger gate samples of both material systems we find an edge resistivity $\rho_\mathrm{edge} = 2\rho_\mathrm{gated}(-4\,\mathrm{V}) = 1.5 - 1.8~\mathrm{k}\Omega/\mu$m. Note that the slope of the fit (black dashed line) indicated in Fig.~2(b) and (d) is not the edge resistivity $\rho_\mathrm{edge}$ (see eq. \ref{ldep}).
% As explained earlier, the plotted resistance $R_{xx}$ is the sum of contributions from the ungated and the gated regions of the Hall bar.

%Proposed \cite{LiuPRL2008} and also measured \cite{QUPRL2015} were double gated structures to tune between metallic, normal, and topologically insulating regime. The initial idea to switch the edge conduction on and off by gating between the different regimes is challenging because of the limited backgate tunability, finite bulk conductivity and the recent discovery of trivial edge conduction in the normal insulating regime \cite{NicheleNJP2016, NguyenPRL2016}. The latter is perfectly in line with the above observations of trivial edge conduction in InAs.  

\begin{figure}
  \includegraphics[width=\columnwidth]{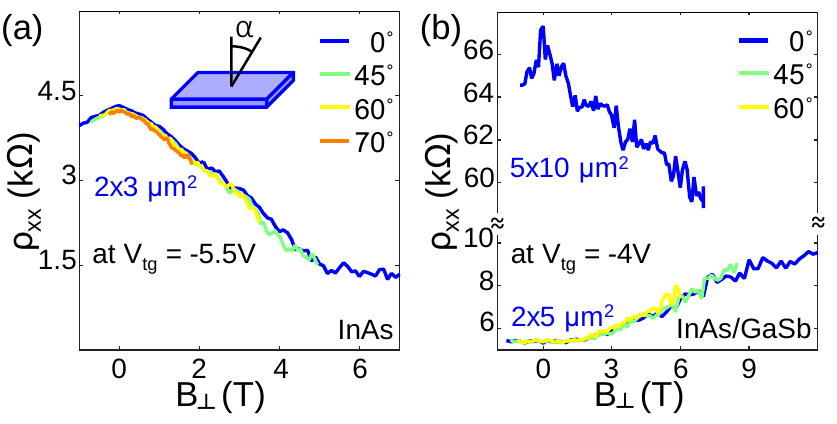}
  \caption{Magnetic field is applied with an angle $\alpha$ to the normal of an InAs two-dimensional electron gas (a) resp. coupled InAs/GaSb systems (b) as schematically explained in the inset of (a). The longitudinal resistivity $\rho_{xx}$ in an edge dominated regime (at a gate voltage of $V_\mathrm{tg(i)} = -5.5\,$V resp. $V_\mathrm{tg(i)} = -4\,$V) is measured on standard Hall bar samples and plotted versus the perpendicular component of the field $B_\mathrm{\bot}$.}
\label{magfield}
\end{figure}

\section{Transport in tilted magnetic fields}

In order to obtain information about the spatial extent of orbital states in the conducting channels along the edges we measure the resistance of micron-sized Hall bars in magnetic fields tilted by an angle $\alpha$ with respect to the normal of the sample plane [inset of Fig.~\ref{magfield}(a)]. Figure~\ref{magfield}(a) displays the resistivity of an InAs sample (wafer A, Hall bar width $2\,\mu\mathrm{m}$, length $3\,\mu$m) measured at $V_\mathrm{tg}=-5.5$\,V, i.e., in the regime dominated by edge conductance, as a function of the magnetic field component $B_\perp$ for various angles $\alpha$. We find that all curves scale on top of each other when plotted against $B_\mathrm{\perp}$. This suggests an anisotropy of the conducting regions at the edges with a larger extent in the plane of the sample than normal to it. Furthermore, the resistance decreases with increasing magnetic field. Theories describing such a trend in narrow channels of two-dimensional systems are found, for example, in Ref.~\onlinecite{BEENAKKE19911}.

The measurements shown in Fig.~\ref{magfield}(b) were obtained on InAs/GaSb (wafer C, Hall bar width $2\,\mu\mathrm{m}$, length $5\,\mu$m and $5\,\mu\mathrm{m}\times10\,\mu\mathrm{m}$, respectively), again in the regime dominated by edge conduction at $V_\mathrm{tg}=-4$\,V (c.f. Fig.~1 in Ref.~\onlinecite{MuellerPRB2015}). Resistivities taken at different tilt angles scale on top of each other when plotted versus $B_\perp$, like in InAs.
Again, we interpret this finding with a larger extent of the conducting edge-region within the plane of the sample than in growth direction. 
%The findings are summarized for InAs (wafer A, $2\times3\,\mu\mathrm{m}^{2}$ Hall bar) in Fig.~3(a).

% and for InAs/GaSb (wafer C, mesoscopic $2\times5\,\mu\mathrm{m}^{2}$ and intermediate $5\times10\,\mu\mathrm{m}^{2}$ Hall bar) in Fig.~3(b). For InAs devices, the resistivity in the edge dominated regime ($V_\mathrm{tg(i)} = -5.5\,$V for InAs devices, see Fig.~1 and $V_\mathrm{tg(i)} = -4\,$V for InAs/GaSb devices, see Fig.~1 in reference \cite{MuellerPRB2015}) was measured as a function of magnetic field applied with a specific angle $\alpha$ to the normal of the sample plane (schematic illustration in the inset of Fig.~3(a)). All measurement were taken at 1.5\,K, except the mesoscopic InAs/GaSb device which was measured at 125\,mK, affirming within 5\% the results obtained at 1.5\,K. For each sample all measurements collapse onto the same curve, when plotted against the perpendicular field component $B_\mathrm{\bot}$. This suggests an anisotropy of the conducting regions at the edges with a larger extent in the 2DEG plane than normal to it. 

%The edge resistivity of InAs devices reduces with increasing perpendicular magnetic field (for all measured device sizes), similar to suppressed back scattering in narrow 2DEG channels \cite{BEENAKKE19911}. 

In InAs/GaSb the trend of the edge resistivity with increasing field depends on the Hall bar size. The edge resistivity of the larger device is suppressed with field like for InAs Hall bars. For small devices the trend in magnetic field is opposite. The reason for such a dependence on device size remains to be explained.

For InAs/GaSb samples in the inverted regime Du {\it et al} \cite{DuPRL2015} also find an increasing edge resistance with magnetic field for small four-terminal devices. Nichele {\it et al} \cite{PRLNichel2014} showed an increase in resistivity with rising field for large devices, which is not in agreement with the findings here, but due to large bulk conductivity their measurement is not in an edge dominated regime. The same applies to magnetic field dependent measurements in Ref.~\onlinecite{PRBKaralic2016}. The decrease in resistance around charge neutrality was attributed to the enhanced anisotropy of the band structure in parallel field. The latter could not be observed for the disordered material presented here. The magnetic field dependence on InAs/GaSb samples in the trivial regime is measured in Ref.~\onlinecite{NguyenPRL2016}, but hard to extract from color plots. It can therefore not be compared with the results presented here.

%Two different trends for the two regimes of sample size open the possibility for two device regimes with potentially different physics. At this point we would like to stress, that the bulk in InAs devices can be depleted completely allowing to study the edge properties detached. The bulk of InAs/GaSb can not fully be depleted and might be magnetic field dependent as well. But it can be expected that the bulk effects are independent on device size. Therefore, the different trends in perpendicular magnetic field for mesoscopic and intermediate devices can be attributed to edge physics.

\section{Critical discussion of edge conduction in $\mbox{InAs/GaSb}$ double quantum wells.}

\begin{figure}
  \includegraphics[width=\columnwidth]{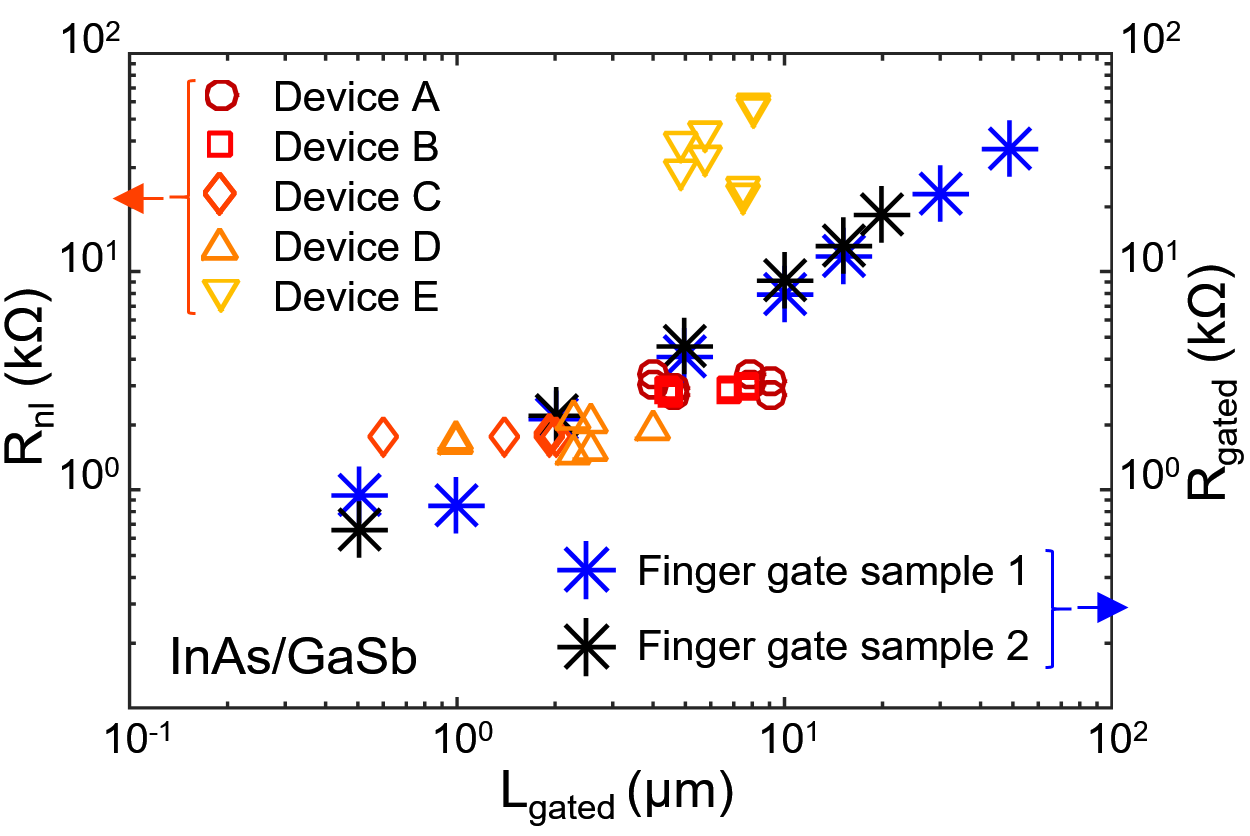}
  \caption{Summary of InAs/GaSb devices from Fig.~\ref{finger}(d) in this paper, and from Ref.~\onlinecite{MuellerPRB2015} measured in the edge dominated regime. The black and blue stars represent the calculated edge resistances $\mathrm{R_{gated}}$ (derivation explained in text) versus the respective gate length $\mathrm{L_{gated}}$ of the finger-gate sample in Fig.~\ref{finger}(d). The other symbols refer to the Hall bar devices presented in Ref.~\onlinecite{MuellerPRB2015}. Plotted is the four terminal non-local resistance $\mathrm{R_{nl}}$ against the gated edge between the respective pair of voltage probes (details in text).}
  \label{summary}
\end{figure}

Based on the new insights obtained from the measurements presented in this paper we now try to critically discuss edge conduction and its length-dependence in InAs/GaSb devices. In this endeavor, we take the data presented in Fig.~\ref{finger}(d) and data from Ref.~\onlinecite{MuellerPRB2015} into account, which were all measured on devices from the same wafer C. Figure~\ref{summary} presents a summary of all the data measured in our lab. The blue and black stars in this figure represent the data points from Fig.~\ref{finger}(d). Plotted is the resistance $R_\mathrm{gated}=2\rho_\mathrm{gated}(-4\,\mathrm{V})L_\mathrm{tg}^{(i)}$ [c.f. Eq.~\eqref{ldep}] versus the gate length $L_\mathrm{tg}^{(i)}$. All the other colored symbols are non-local resistances $R_\mathrm{nl}$ from the different devices of Ref.~\onlinecite{MuellerPRB2015} (in Ref.~\onlinecite{MuellerPRB2015} referred as truly non-local resistances of type 1). In Fig.~3(a) of this reference, only the data points of device C were explicitly shown , but all devices were analyzed and it was concluded that the edge resistance is {\em independent} of edge length, in apparent contrast to the data in Fig.~\ref{finger}(d).

First we note that the resistance $R_\mathrm{gated}$ in Fig.~\ref{summary} is only a {\em lower bound} for the true edge resistance. The measurements on the finger gate sample do not distinguish bulk- and edge contributions to the total current. If there was a bulk current in the InAs/GaSb finger gate devices, then $R_\mathrm{gated}$ would underestimate the true edge resistance.
 
The resistances for devices A--E from Ref.~\onlinecite{MuellerPRB2015} plotted in Fig.~\ref{summary} are bare non-local resistances $R_\mathrm{nl}^{(i)}$ obtained by dividing particular measured non-local voltages $V_\mathrm{nl}^{(i)}$ by $I_\mathrm{tot}$, the total current applied. These resistances are plotted in Fig.~\ref{summary} against the lengths $L_\mathrm{gated}^{(i)}$ of the gated edge $i$ between the respective pair of voltage probes. Here, the striking phenomenon is that the bare non-local resistance is independent of $i$, and therefore of $L_\mathrm{gated}^{(i)}$. For this reason it was concluded in Ref.~\onlinecite{MuellerPRB2015} that the edge transport is ballistic on the investigated length scales. We see in Fig.~\ref{summary} that this holds true for the small devices A--D, but is no longer found for the larger device E. This behavior of InAs/GaSb is in stark contrast to the non-local resistances of the investigated InAs devices [c.f. Fig.~1(c)], where we found scaling with $L_\mathrm{gated}^{(i)}$.

However, the $R_\mathrm{nl}^{(i)}$ cannot be directly interpreted as edge resistances, because $I_\mathrm{tot}$ is the sum of two edge currents of possibly unequal magnitude running along the two Hall bar edges of unequal lengths between the current contacts, plus a possible bulk current. Assuming completely ballistic edge conduction and zero bulk current we find that the edge resistance is larger than $R_\mathrm{nl}^{(i)}$ by a factor of two. Diffusive edge conduction and also a finite bulk current would raise this factor further. This would move all data points of devices A--E above the lower bound of the edge resistance given by the black and blue stars of the finger gate sample in Fig.~\ref{summary}. Based on these considerations, we may state that the edge resistances of all the investigated devices are of the same order of magnitude and consistent with each other.

The question still remains, why the non-local measurements on Hall bars give length independent $R_\mathrm{nl}^{(i)}$, whereas the finger gate sample exhibits a linear length dependence down to at least $1\,\mu$m. In order to find possible sources of misinterpretations here, we take a critical look at the edge lengths $L_\mathrm{gated}^{(i)}$ extracted for the devices in Ref.~\onlinecite{MuellerPRB2015}. These lengths were taken from optical microscope images assuming that the width of conductive edge regions is much smaller than any lithographic width of the samples. However, considering the finding of a finite extent of the edge conducting regions in the plane of the sample (c.f. Fig.~\ref{magfield}) , it is conceivable that the conducting regions cannot enter the narrow voltage probes without coupling so strongly that the gated edge length within these voltage probes does not contribute to the relevant edge length. This scenario would reduce the spread of the true $L_\mathrm{gated}^{(i)}$ in Fig.~\ref{summary} so strongly that a length-independent resistance could no longer be deduced from the data with sufficient confidence.

Similarly, one could find reasons why the linear dependence of $R_\mathrm{gated}$ on gate length $L_\mathrm{tg}^{(i)}$ arises in spite of the presence of helical edge modes in the finger-gate sample of this paper. One possible scenario is the presence of bulk or trivial edge conductance shunting the significantly lower conductance of the helical edge modes, which is expected to be $e^2/h$.

Summarizing this critical discussion of our own measurements, we have to state that, first, the linear length dependence in the finger-gate sample only gives a lower bound for the possible edge resistances which is well below the value expected for helical edge states at least up to lengths of $10\,\mu$m. Second, the non-local resistances of Hall-bar devices do not give a robust estimate of edge resistances either, because the relevant current along a particular sample edge is not known. Third, the edge-length estimates for these samples are based on the assumption of narrow edge channels  that may well be violated, which would render the length-independent edge resistance an illusion. We believe that this discussion bears importance also beyond our data for the interpretation of related work on transport in InAs/GaSb double quantum wells by other authors.

\section{Conclusion}

Our experiments show edge conduction in InAs two-dimensional electron gases where no topological effects are expected. An edge resistivity of $\rho_\mathrm{edge} = 1.3-2.5\,\mathrm{k}\Omega/\mu$m could be confirmed for standard as well as asymmetric Hall bars and finger gate samples. These results have to be compared to investigations in the InAs/GaSb double quantum well system, a topological insulator candidate. The latter also shows a resistance with linear dependence on edge length of the same order of magnitude for edge lengths as small as $1\,\mu$m. Additionally, both systems show a magnetoresistance in tilted field that is independent of the parallel magnetic field component with respect to the sample plane. Even though standard InAs/GaSb samples have indications for edge length independent non-local resistances, an alternative, trivial explanation can not be excluded with the latter results in mind. The presented investigations motivate us to optimize sample processing in order to suppress trivial edge conduction or to enhance the spin-relaxation length of the topological edges. The precise length dependence of the trivial edge conduction could be an important aspect in view of the clearcut identification of the QSH phase in InAs/GaSb systems. 

%exceeding $\mathrm{l}>5\,\mu$m. For small Hall bar devices this report could summarize three different properties of the edge resistivity of the two material systems: A different type of hysteresis, a different scaling behavior on edge length and a different trend in perpendicular field. The latter could be an important aspect towards the clearcut identification of the QSH phase in InAs/GaSb systems. 

The authors thank the Swiss National Science Foundation for financial support via NCCR QSIT (Quantum Science and Technology).

%\begin{table*}
	%\centering
		%\begin{tabular}{|l|l|l|l|l|l|}
			%\hline
			%\textbf{system} & \textbf{mobility} & \textbf{HB size} & \textbf{$R_{xx}(B_\mathrm{\bot})$} & \textbf{$R_{xx}(B_\mathrm{\parallel})$} & comment\\
			%\hline
			%\hline
			%InAs & high & large & $\searrow$ & unknown & finger gate sample $50~\mu m$ gate\\
			 %&  & small & $\searrow$ & $\longrightarrow$ & this paper\\
			%\hline
			%InAs/GaSb & high & large & $\nearrow$ & $\searrow$ & Fabrizio\\ 
			 %&  & small & $\nearrow$ & unknown & \\
			 %& low & large & $\searrow$ & unknown & Device E\\
			 %&  & small & $\nearrow$ & $\longrightarrow$ & this paper\\
			%\hline
		%\end{tabular}
		%\caption{...}
%\end{table*}

\bibliography{edgesInAscite}
	
\end{document}